\documentclass[aps,prd,10pt,onecolumn,nofootinbib,showpacs,showkeys,notitlepage]{revtex4-1}
\usepackage{graphicx,bm}
\usepackage{hyperref}
\usepackage{amssymb}

\newcommand{\eqref}[1]{(\ref{#1})}

\renewcommand{\d}{\partial}

\newcommand{\nn}{\nonumber\\}

\newcommand{\x}{{\bf x}}

\newcommand{\rh}{\varrho}

\newcommand{\exv}[1]{\left\langle{#1}\right\rangle}

\newcommand{\ep}{\varepsilon}

\newcommand{\sgn}{\mathop{\textrm{sgn}}}

\newcommand{\pint}[2]{{\int\!\frac{d^{#1}#2}{(2\pi)^#1}\,}}
\newcommand{\pintz}[1]{{\int\!\frac{d #1}{2\pi}\,}}

\newcommand{\G}{{\cal G}}
\newcommand{\bG}{\bar{\cal G}}
\newcommand{\brh}{\bar\rh}
\newcommand{\bz}{\bar\zeta}
\newcommand{\JJ}{{\cal J}}
\newcommand{\bJJ}{\bar{\cal J}}

\begin{document}

\title{Spectral function of the Bloch-Nordsieck model at finite temperature}

\author{A. Jakov\'ac\hspace*{0.2em}} \email{jakovac@phy.bme.hu}
\affiliation{Institute of Physics, E\"otv\"os University, H-1117
  P\'azm\'any P\'eter s\'et\'any 1/A, Budapest, Hungary}
\author{P. Mati\hspace*{0.2em}} \email{mati@phy.bme.hu} 
\affiliation{Institute of Physics, Budapest University of Technology
  and Economics, H-1111 Budafoki ut 8, Budapest, Hungary}
\date{\today}

\begin{abstract}
  In this paper we determine the exact fermionic spectral function of
  the Bloch-Nordsieck model at finite temperature. Analytic results
  are presented for some special parameters, for other values we have
  numerical results. The spectral function is finite and normalizable
  for any nonzero temperature values. The real time dependence of the
  retarded Green's function is power-like for small times and exhibits
  exponential damping for large times. Treating the temperature as an
  infrared regulator, we can also give a safe interpretation of the
  zero temperature result.
\end{abstract}

\maketitle

\section{Introduction}

The behavior of the ultra-soft regime of massless field theories
presents a serious challenge which, on the other hand, is crucial for
understanding of the most, physically relevant theories. The soft
nature of the excitations perturbatively leads to infrared (IR)
divergences in various physical quantities such as the self-energy
near the mass shell. To have reliable results, one has to resum the
most sensitive part of the IR physics. The identification of the
sources of these divergences, the elaboration of the appropriate
mathematical tools and finally the realisation of the resummation
itself is a formidable task. Moreover, the details can depend on the
environment, that is the resummation in deep inelastic scattering and
at finite temperature equilibrium may require different approaches. It
is not a surprise, therefore, that so many resummation methods exist,
working at different circumstances.


In this sensitive field, where the physical reliability of a
resummation may crucially depend on the correct identification of the
relevant sources of the IR divergences, it is quite valuable to find a
model which is physically motivated and exactly solvable. This is the
reason why so many two-dimensional conformal and integrable theories
have relevance. In four dimensions, however, exactly solvable models
are much rarer.

A physically well-motivated model is the Bloch-Nordsieck (BN) model
\cite{BN}. In its long history it became a text-book material
\cite{BogoljubovShirkov,Fried}. Physically it corresponds to the deep
IR limit of QED, where the photons have no energy even for a fermion
spin-flip. In particular it can be used to prove QED theorems in this
energy regime \cite{Weldon:1991eg}. The BN model can be solved
exactly, the photon contributions can be fully summed up. The fermion
propagator has been calculated at zero temperature both based on
functional methods \cite{BogoljubovShirkov,Fried}, and with help of
Dyson-Schwinger equations \cite{Alekseev:1982dk,Jakovac:2011aa}, where
also a detailed renormalization analysis is possible. The spectral
function of the model at zero temperature reads in Feynman gauge as
\begin{equation}
  \rh(w,T=0) \sim w^{-1-\frac\alpha\pi}, \qquad\mathrm{where}\quad
  w=u_\mu p^\mu-m,\quad  \alpha=\frac{e^2}{4\pi}.
\end{equation}
Here $u$ is a 4-vector parameter of the model, loosely identifiable
with the four-velocity of the fermion. This function, however, has a
singular behavior: it is not normalizable, therefore the sum rule
$\int \rho=2\pi$ can be satisfied only with zero wave function
renormalization factor. Moreover, the naive inverse Fourier transform
of this function is $\sim t^{\alpha/\pi}$ describing growth of
correlation in time. The correct physical interpretation of these
results requires some IR regulator, which can be, for example, the
temperature.

At finite temperature the model is much less studied. In the seminal
papers of Blaizot and Iancu \cite{IancuBlaizot1,IancuBlaizot2} the
authors studied the large time behavior of the fermion propagator with
the Hard Thermal Loop (HTL) improved photon propagator. Using this
result, Weldon worked out a spectral function which is valid in the
vicinity of the mass shell \cite{Weldon:2003wp}. With a different
approach, Fried \emph{et. al.} studied the time dependence of the
momentum loss of a hard incoming fermionic particle
\cite{Fried:2008tb}.

We have several goals in this paper. The main goal is to work out the
complete spectral function of the BN model for all momenta, and see
how the short time dynamics, resembling the $T\to 0$ limit, goes over
to the long time damping. Because of the relative simplicity of the
model we can even give analytic solutions for certain parameters,
while for other, analytically not reachable parameter values we used a
well controlled numerical procedure. Another goal is to extend our
Dyson-Schwinger formalism combined with Ward identities
\cite{Jakovac:2011aa}, which works excellently at zero temperature, to
finite temperatures. With the help of it, the complete renormalization
process remains fully controlled.

Our paper will be organized as follows. First, we define the
Bloch-Nordsieck model in Section \ref{sec:BNdef}. We review the
Dyson-Schwinger equations and Ward identities in finite temperature
real time formalism, and apply them to the Bloch-Nordsieck model. In
Section \ref{sec:solve} we solve these equations. At zero velocity
(Subsection \ref{sec:u0}) we provide an analytic formula for the
fermion propagator, supported by a numerical verification. At nonzero
velocity (Subsection \ref{sec:ufin}) we solve them numerically. In
Subsection \ref{sec:disc} we compare our results with previous works
in the literature. In Section \ref{sec:concl} we give the conclusions
of the paper.

\section{The Bloch Nordsieck model at finite temperature}
\label{sec:BNdef}

The Bloch-Nordsieck model is the low energy limit of QED, where we
take into account only a single spin orientation. Its Lagrangian is
related to the QED Lagrangian by changing the Dirac matrices
$\gamma^\mu$ for a four-vector $u^\mu$:
\begin{equation}
  {\cal L} = -\frac14 F_{\mu\nu} F^{\mu\nu} + \Psi^\dagger(iu_\mu D^\mu
  - m)\Psi,\qquad iD_\mu = i\d_\mu -eA_\mu,\quad F_{\mu\nu}=\d_\mu
  A_\nu - \d_\nu A_\mu.
\end{equation}
We can choose $u$ to be a four-velocity, or it can be
$u=(1,\mathbf{v})$: the two are related by a simple field and mass
rescaling, since by replacing $\Psi\to \Psi/\sqrt{u_0}$ and $m\to m
u_0$, we can reach the $u_0=1$ scenario. The quantity $\mathbf{v}
=\mathbf{u}/u_0$ can be interpreted as the velocity of the fermion. 

We are interested in the finite temperature fermion propagator. To
determine it, we use the real time formalism (for details, see
\cite{LeBellac}). Here the time variable runs over a contour
containing forward and backward running sections ($C_1$ and
$C_2$). The propagators are subject to boundary conditions which can
be expressed as the KMS (Kubo-Martin-Schwinger) relations. The
physical time can be expressed through the contour time $t={\cal
  T}(\tau)$. This makes possible to work with fields living on a
definite branch of the contour, $\Psi_a(t,\x) = \Psi(\tau_a,\x)$ where
${\cal T}(\tau_a)=t$, and $\tau_a\in C_a$ for $a=1,2$; and similarly
for the gauge fields. The propagators are matrices in this notation:
\begin{equation}
  i{\cal G}_{ab}(x)=\exv{T_C \Psi_a(x) \Psi_b^\dagger (0)}\qquad
  \mathrm{and}\qquad iG_{\mu\nu,ab}(x)=\exv{T_C A_{\mu a}(x) A_{\nu b} (0)},
\end{equation}
where $T_C$ denotes ordering with respect to the contour variable
(contour time ordering). $G_{11}$ corresponds to the Feynman
propagator, and, since the $C_2$ contour times are always larger than
the $C_1$ contour times, $G_{21}=G^>$ and $G_{12}=G^<$ are the
Wightman functions. The KMS relation for a bosonic/fermionic
propagator reads $G_{12}(t,\x)= \pm G_{21}(t-i\beta,\x)$ which has the
following solution in Fourier space
\begin{equation}
  \label{eq:id1}
  iG_{12}(k) = \pm n_\pm(k_0) \rh(k),\qquad iG_{21}(k) = (1\pm
  n_\pm)(k_0) \rh(k),
\end{equation}
where 
\begin{equation}
  n_\pm(k_0) = \frac1{e^{\beta k_0}\mp 1}\quad \mathrm{and}\quad \rh(k)
  = iG_{21}(k)-iG_{12}(k)
\end{equation}
are the distribution functions (Bose-Einstein (+) and Fermi-Dirac (-)
statistics), and the spectral function, respectively. It is sometimes
advantageous to change to the R/A formalism with field assignment
$\Psi_{1,2} = \Psi_r\pm\Psi_a/2$. Then one has $G_{aa}=0$ for both the
fermion and the photon propagators. The relation between the $1\,2$
and the R/A propagators reads
\begin{equation}
  \label{eq:id2}
  G_{rr}=\frac{G_{21}+G_{12}}2,\quad G_{11} = G_{ra} + G_{12},\quad
  \rh=i G_{ra}-iG_{ar}.
\end{equation}
The $G_{ra}$ propagator is the retarded, the $G_{ar}$ is the advanced
propagator, $G_{rr}$ is usually called the Keldysh propagator.

At zero temperature the fermionic Feynman-propagator reads:
\begin{equation}
  {\cal G}_0(p) = \frac1{u_\mu p^\mu -m +i\ep}.
\end{equation}
It has a single pole which means that there is no antiparticles in the
model. Consequently, all closed fermion loops are zero, thus there is
no self-energy correction to the photon propagator at zero
temperature. Physically this means that the energy is not enough to
excite the antiparticles. In fact, if we interpret the $u$ parameter
as the four-velocity of the fermion, the Bloch-Nordsieck model
describes that regime where the soft photon fields do not have energy
even for changing the velocity of the fermion (no fermion
recoil). This leads to the interpretation that the fermion is a hard
probe of the soft photon fields, and as such it is not part of the
thermal medium \cite{IancuBlaizot2}. So we will set ${\cal G}_{12}=0$,
therefore the closed fermion loops as well as the photon self energy
remain zero even at finite temperature. Another, mathematical reason,
why we must not consider dynamical fermions -- which could show up in
fermion loops -- is that the spin-statistics theorem
\cite{PeshkinSchroeder} forbids a one-component dynamical fermion
field.

This means that now the \emph{exact} photon propagator reads in
Feynman gauge
\begin{equation}
  G_{ab,\mu\nu}(k) = -g_{\mu\nu} G_{ab}(k),\qquad G_{ra} =
  \frac1{k^2}\biggr|_{k_0\to k_0+i\ep},\quad \rh(k)= 2\pi\sgn(k_0) \delta(k^2),
\end{equation}
all other propagators can be expressed using identities \eqref{eq:id1}
and \eqref{eq:id2}.

\subsection{Dyson-Schwinger equations}

The operator equations of motion give relations of the different
Green's functions, formulated as the Dyson-Schwinger equations. These
equations are \emph{local}, and so they are valid in generic
non-equilibrium situations, and, of course, in a thermal medium, too.

The generating form of the Dyson-Schwinger equations for generic
fields $\Phi_i$ reads \cite{Collins}
\begin{equation}
  \label{eq:SDgen}
  \exv{\frac{\delta S}{\delta \Phi_i(y)} \Phi_{a_1}(x_1)\dots
    \Phi_{a_n}(x_n)} = i \sum\limits_{k=1}^n
  \delta_{ia_k}\delta(y-x_k) \exv{\Phi_{a_1}(x_1)\dots
    \Phi_{a_{k-1}}(x_{k-1})\Phi_{a_{k+1}}(x_{k+1})\dots \Phi_{a_n}(x_n)}.
\end{equation}
In real time formalism the time variable is the contour time (usually
it is the variable of the path integral). We define the fermionic self
energy in the usual way
\begin{equation}
  {\cal G}(x,y) ={\cal G}^{(0)}(x,y) + \int_C\! d^4x' d^4y'\, {\cal
    G}^{(0)}(x,x') \Sigma(x',y'){\cal G}(y',y),
\end{equation}
where the symbol $\int_C$ means time integration over the
contour. Then we find in the Bloch-Nordsieck model
\begin{equation}
  \Sigma(x,y) = i\alpha(x_0) e^2u_\mu \int_C\! d^4w d^4z\, {\cal G}(x,w)
  G^{\mu\nu}(x,z) \Gamma_\nu(z;w,y),
\end{equation}
where the tree level vertex is $eu_\mu$, the proper vertex is denoted
by $e\Gamma_\mu$, and $\alpha(x_0)$ is 1 if $x_0\in C_1$ and $-1$ if
$x_0\in C_2$. This factor appears because we expressed the functional
derivative $\frac{\delta S}{\delta \Phi_i(y)}$ through the derivatives
of the Lagrangian, which, however, changes sign on $C_2$.

\begin{figure}[htbp]
  \centering
  \includegraphics[width=0.3\textwidth]{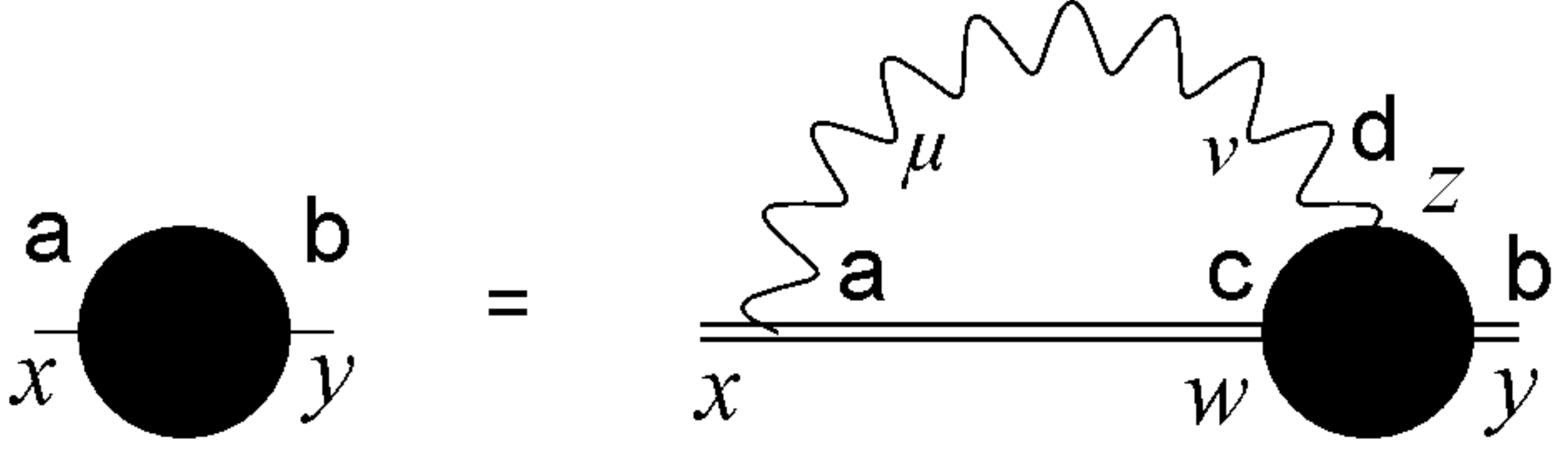}
  \caption{The Dyson-Schwinger equations in real time formalism}
  \label{fig:DS}
\end{figure}

We can also express this equation with the two-component notation as
it can be seen on Fig.~\ref{fig:DS}.
In terms of analytic formulas it reads:
\begin{equation}
  \Sigma_{ab}(x,y) = i\alpha_a e^2u_\mu \sum_{c,d=1}^2 \int\! d^4w
  d^4z\, {\cal G}_{ac}(x,w) G^{\mu\nu}_{ad}(x,z) \Gamma_{\nu;dcb}(z;w,y),
\end{equation}
where $\alpha_a=(-1)^{a+1}$. In Fourier space it reads:
\begin{equation}
  \Sigma_{ab}(p) = i\alpha_a e^2u_\mu \sum_{c,d=1}^2 \pint 4k {\cal
    G}_{ac}(p-k) G^{\mu\nu}_{ad}(k) \Gamma_{\nu;dcb}(k;p-k,p).
\end{equation}

\subsection{The vertex function in the Bloch-Nordsieck model}

The second use of the Dyson-Schwinger equation is to have a form for
the vertex function. From \eqref{eq:SDgen} we find for any gauge
theories
\begin{equation}
  \exv{\frac{\delta S}{\delta A^\mu(x)} O(\bar\Psi,\Psi)} =0,
\end{equation}
where $O$ is any local operator containing $\bar\Psi$ and
$\Psi$. This implies, in particular
\begin{equation}
  \exv{A_\mu(x) \Psi(y) \bar\Psi(z)} = \int_C d^4x'
  G_{\mu\nu}(x,x') \exv{ j^\nu(x') \Psi(y)\bar\Psi(z)},
\end{equation}
where $j_\mu$ is the conserved current. The vertex function shows up
in the $A\Psi\Psi^\dagger$ correlator as
\begin{equation}
  \exv{A_\mu(x)\Psi(y)\bar\Psi(z)} = \int_C\! d^4x' d^4y'd^4z'\,
  iG_{\mu\nu}(x,x') i{\cal G}(y,y')( -ie) \Gamma^\nu(x',y',z') i{\cal
    G}(z',z).
\end{equation}
From here we find
\begin{equation}
  \label{eq:vfunandj}
  \int_C d^4y'd^4z' \,i{\cal G}(y,u)\, e\Gamma^\mu(x;u,v) i{\cal
    G}(v,z) = \exv{j^\mu(x) \Psi(y) \bar\Psi(z)}.
\end{equation}
In the BN model the fermion propagator is a scalar, moreover $j_\mu =
eu_\mu \Psi^\dagger\Psi$ is proportional to $u_\mu$. Therefore the
vertex function is proportional to $u^\mu$, too. This is written in
the Fourier space as
\begin{equation}
  \label{eq:vertexu}
  \Gamma^\mu(k;p,q) = u^\mu \Gamma(k;p,q)\, (2\pi)^4 \delta(k+p-q),
\end{equation}
where we also used the energy-momentum conservation.

\subsection{Ward identities}

The local equations expressing current conservation can be used in a
similar manner. The generating form reads
\begin{equation}
  \frac{\d}{\d x^\mu} \exv{j^\mu(x) \Phi_{a_1}(x_1)\dots
    \Phi_{a_n}(x_n)} = -i \sum\limits_{k=1}^n
  \delta_{ia_k}\delta(x-x_k) \exv{\Phi_{a_1}(x_1)\dots
    \Phi_{a_{k-1}}(x_{k-1})\Delta \Phi_i(y) \Phi_{a_{k+1}}(x_{k+1})\dots
    \Phi_{a_n}(x_n)},
\end{equation}
where $\Delta \Phi_i$ is the transformation of the $i$th field
generated by the conserved charge $Q=\int d^3\x j^0(t,\x)$. This
means, in particular
\begin{equation}
  \frac{\d}{\d x^\mu} \exv{j^\mu(x) \Psi(y)\bar\Psi(z)} = e
  \delta(x-z) {\cal G}(y,z) - e \delta(x-y) {\cal G}(y,z).
\end{equation}
We can write the corresponding equation for the vertex function, using
\eqref{eq:vfunandj}:
\begin{equation}
  \frac{\d}{\d x^\mu} \int_C d^4ud^4v\, i{\cal G}(y,u)
  \Gamma^\mu(x;u,v) i{\cal G}(v,z) =  \delta(x-z) {\cal G}(y-z) -
  \delta(x-y) {\cal G}(y-z).
\end{equation}
This form is easy to rewrite in the two-component formalism, taking
into account that to satisfy the delta function the time arguments
must be on the same contour. One finds in Fourier space
\begin{equation}
  k_\mu \Gamma^\mu_{abc}(k;p,q) = \left[ \delta_{ab} {\cal
      G}_{bc}^{-1}(q) - \delta_{ac} {\cal G}_{bc}^{-1}(p) \right] \,
  (2\pi)^4 \delta(k+p-q).
\end{equation}

In the Bloch-Nordsieck model, because of the special property of the
vertex function expressed in \eqref{eq:vertexu}, the vertex function
is \emph{completely} determined by the fermion propagator:
\begin{equation}
  \label{eq:vertex}
  \Gamma_{abc}(k;p,q) = \frac1{uk}\left[  \delta_{ab} {\cal
      G}_{bc}^{-1}(q) - \delta_{ac} {\cal G}_{bc}^{-1}(p)\right]
  \biggr|_{p=q-k}.
\end{equation}

\section{Solution of the Dyson-Schwinger equations}
\label{sec:solve}

Since the vertex function in the Bloch-Nordsieck model can be
expressed with the fermion propagator, the Dyson-Schwinger equations
for the fermion propagator become closed. At zero temperature it can
be shown \cite{Alekseev:1982dk,Jakovac:2011aa} that the solution of
this equation yields the same result as the functional techniques,
moreover, renormalization can be fully controlled here.  In this
Section we discuss the solution at finite temperature.

We will use Feynman gauge, and denote the photon propagator as
$G_{\mu\nu}=-g_{\mu\nu} G$. Then this closed equation can be written
as
\begin{eqnarray}
  \Sigma_{ac}(p) &&= -ie^2U^2 \alpha_a \sum_{a',b'=1}^2\pint4k
  \frac1{uk} G_{aa'}(k) \G_{ab'}(p-k) \biggl[ \delta_{a'b'}
  (\G^{-1})_{b'c}(p) - \delta_{a'c} (\G^{-1})_{b'c}(p-k)\biggr]
  =\nn&&= -ie^2U^2 \alpha_a \left[\sum_{a'=1}^2 (\G^{-1})_{a'c}(p)
    \pint4k \frac1{uk} G_{aa'}(k) \G_{aa'}(p-k) -
    \delta_{ac} \pint4k \frac1{uk} G_{aa}(k)\right],
\end{eqnarray}
where $U^2=u_0^2-\mathbf{u}^2$. In particular
\begin{eqnarray}
  \label{eq:12formalismsigma}
  \Sigma_{11}(p) &&= -ie^2U^2 \left[ \sum_{a'=1}^2 (\G^{-1})_{a'1}(p)
    \pint4k \frac1{uk} G_{1a'}(k) \G_{1a'}(p-k) - \pint4k
    \frac{G_{11}(k)}{uk} \right]\nn
  \Sigma_{12}(p) &&= -ie^2U^2 \sum_{a'=1}^2 (\G^{-1})_{a'2}(p) \pint4k
  \frac1{uk} G_{1a'}(k) \G_{1a'}(p-k).
\end{eqnarray}

Instead of $G_{11}$ and $G_{22}$ it is more aesthetic to work with the
retarded and advanced propagators (the relations are given in
\eqref{eq:id2}). Since in the R/A formalism $G_{aa}=0$, the retarded
propagator satisfies a homogeneous self-energy relation
\begin{equation}
  G_{ra}(p) = G_{ra}^{(0)}(p) + G_{ra}^{(0)}(p) \Sigma_{ar}(p) G_{ra}(p),
\end{equation}
while the propagators in the $1,2$ components mix. From the
definitions we easily find
\begin{equation}
  \Sigma_{ar} = \Sigma_{11}+\Sigma_{12},\qquad \G_{11} - \G_{12} = \G_{ar}.
\end{equation}
Therefore we have, using \eqref{eq:12formalismsigma} and \eqref{eq:id2}
\begin{equation}
  \label{eq:sigmaandJ}
  \Sigma_{ar}(p) = \JJ(p) \G^{-1}_{ra}(p) - \Delta M,
\end{equation}
where
\begin{eqnarray}
  \label{eq:JJ1}
  && \JJ(p) = -ie^2U^2 \pint4k \frac1{uk} \left(G_{21}(k) \G_{ra}(p-k) -
    G_{ra}(k) \G_{12}(p-k) \right),\nn
  && \Delta M = -ie^2U^2 \pint4k \frac{G_{11}(k)}{uk}.
\end{eqnarray}

It is easy to see that $\Delta M=0$. The $11$ photon propagator
$G_{11}$ is even for $k\to-k$, which is true in general, but now we
can prove by inspecting the free propagator which is exact in our case
\begin{equation}
  iG_{11}(k) = \frac{i}{k^2+i\ep} + (n(k_0)+\Theta(-k_0))
  2\pi\sgn(k_0)\, \delta(k^2).
\end{equation}
For the first term the $k\to-k$ symmetry is evident, in the second we
should use the identity $n(k_0) + n(-k_0)+1=0$. Therefore with the
change $k\to-k$ of the integration variable, the $G_{11}$ propagator
remains the same while $uk$ changes sign, so $\Delta M$ changes sign,
too. As a consequence $\Delta M=0$.

The Bloch-Nordsieck model, as all 4D interacting quantum field
theories, contains divergences. To obtain finite result, we need wave
function, mass and coupling constant renormalization. Since the above
expressions have contained the original parameters of the Lagrangian,
we should rewrite them in terms of the renormalized quantities. From
now on the parameters $m$ and $e$ will denote the renormalized ones,
while $m_0$ and $e_0$ are the bare quantities. Renormalization goes
like in the zero temperature case \cite{Jakovac:2011aa}: assuming that
the renormalized mass $m=Zm_0$ where $Z$ is the fermion wave function
renormalization constant (this is ensured by the Ward identities) we
can write $\G^{-1}_{ra} = Z(up-m) -\Sigma_{ar}$, and from
\eqref{eq:sigmaandJ} we find
\begin{equation}
  \label{eq:master}
  \G_{ra}(p) = \frac{\zeta(p)}{up-m},\qquad\mathrm{where}\qquad
  \zeta(p)=\frac{1+\JJ(p)}Z. 
\end{equation}

\subsection{Calculation of $\JJ$}
\label{sec:zeta}

In the expression of $\JJ$ in eq. \eqref{eq:JJ1} there appears
$\G_{12}(k)$. As we discussed earlier, for the sake of physical and
mathematical consistency of the model, we must assume that the fermion
describes a hard probe, itself is not a dynamical field, which means
that we must set $\G_{12}(k)=0$. Then from \eqref{eq:JJ1} we can
easily recover the zero temperature result \cite{Jakovac:2011aa}. At
finite temperature we have
\begin{equation}
  \label{eq:JJ}
  \JJ(p) = -ie^2U^2 \pint4k \frac1{uk} G_{21}(k) \G_{ra}(p-k).
\end{equation}

Next we prove by recursion that the solution for $\G_{ra}$ depends
solely on $w=up-m$. It is true at tree level where $\G^{-1}_{ra}=
up-m$. So let us assume that $\G_{ra}(p)=\bG_{ra}(up-m)$. Then
\begin{equation}
  \JJ(p) = -ie^2U^2 \pint4k \frac1{uk} G_{21}(k) \bG_{ra}(up-m-uk),
\end{equation}
implying $\JJ(p)=\bJJ(up-m)$. Equation \eqref{eq:master} tells us that
if $\JJ$ depends only on $up-m$, then $\G_{ra}$ also depends only on
$up-m$. With this statement the recursion is closed.

Since in the BN model the free photon propagator is exact, we shall
write it into eq. \eqref{eq:JJ}. Using \eqref{eq:id1} for the $G_{21}$
propagator, and applying the Landau prescription ($w\to w+i\ep$) we
find
\begin{equation}
  \label{eq:JJ2}
  \bJJ(w) = e^2U^2 \pint4k \frac1{uk}\, (1+n(k_0)) \frac{2\pi}{2k}
  (\delta(k_0-k) - \delta(k_0+k))\, \bG_{ra}(w-uk).
\end{equation}
This result, as we shall show in Section \ref{sec:disc}, is consistent
with the results of \cite{IancuBlaizot1,IancuBlaizot2}.

The $k$ integration can be performed, apart from the single component
$q=ku$. We find after a straightforward calculation:
\begin{equation}
  \label{eq:bJJ}
  \bJJ(w) = \frac{-\alpha}{\pi} \int\limits_{-\infty}^\infty
  \!dq\,f(q,u)\, \bG_{ra}(w-q),
\end{equation}
where $\alpha = e^2/(4\pi)$ and
\begin{equation}
  \label{eq:fqu}
  f(q,u) = \frac{u_0(1-v^2)}{2v}
  \int\limits_{u_0(1-v)}^{u_0(1+v)}\!\frac{ds}{us^2}\, (1+n(\frac qs))
  = \frac{u_0(1-v^2)}{2vq\beta} \ln\frac {e^{\beta q/(u_0(1-v))}-1}
  {e^{\beta q/(u_0(1+v))}-1},
\end{equation}
where $u=u_0(1,\mathbf{v})$ and $v=|\mathbf{v}|$ (i.e. $\mathbf{v}$ is
the velocity $\mathbf{v}=\mathbf{u}/u_0$).

At zero temperature $f(q)=\Theta(q)$. At $v=0$ we find
\begin{equation}
  f(q,\bm u=0) = 1+n(q).
\end{equation}

\subsection{Renormalization}
\label{sec:ren}

In \eqref{eq:bJJ} we find ultraviolet (UV) divergences. From the
expression of $f(q,u)$ (eq. \eqref{eq:fqu}) we see that for large
momenta the thermal distribution functions always decrease
exponentially, thus yielding UV finite result. So all the UV
singularity is in the $T=0$ part, discussed already in
\cite{Jakovac:2011aa}.

To apply the renormalized treatment at finite temperature, we recall
some results from \cite{Jakovac:2011aa}. At $T=0$ \eqref{eq:bJJ} can
be written in spectral representation and with dimensional
regularization as
\begin{equation}
  \bJJ_0(w) = \frac{-\alpha}\pi \int\limits_0^\infty
  \!dq\,\bG_{ra}(w-q) = \frac{\alpha}\pi
  \int\limits_{-\infty}^\infty\! \frac{d w'}{2\pi}\, \brh( w') 
  \int\limits_0^\infty \!dq\,\frac1{q+ w'-w-i\ep} = 
  \frac{\alpha}{\pi} \int\limits_{-\infty}^\infty\!
  \frac{d w'}{2\pi}\, \brh( w') \left[{\cal D}_\ep
    -\ln\frac{ w'-w-i\ep}{\mu} \right],
\end{equation}
where
\begin{equation}
  {\cal D}_\ep =\frac1{2\ep} + \frac12\ln(4\pi) +\frac12 P_{1/2}
\end{equation}
($P_{1/2}=-1.96351$ is the value of the polygamma function with
$0,1/2$ arguments).

As we discussed in \cite{Jakovac:2011aa}, the divergent term is
necessary for the coupling constant and wave function
renormalization. We can write, assuming normalizability of $\brh$
\begin{equation}
  \bz(w) = \frac{1+\bJJ(w)}Z = \frac{\displaystyle
    \frac{4\pi^2}{e_0^2}+ {\cal D}_\ep +\bJJ_{fin}(w)}{\displaystyle
    \frac{4\pi^2Z}{e_0^2}},
\end{equation}
where $\bJJ_{fin}(w)$ is finite. We introduce
\begin{equation}
  \frac{4\pi^2}{e_0^2} + {\cal D}_\ep= \frac{4\pi^2}{e^2},\qquad
  \frac{4\pi^2Z}{e_0^2} = \frac{4\pi^2z_r}{e^2},
\end{equation}
where $z_r$ and $e$ now are finite (renormalized) values. Using
renormalization group invariance we can write for the complete finite
temperature contribution
\begin{equation}
  \bz(w) = \frac{\bar e^2}{4\pi^2}
  \left[\int\limits_{-\infty}^\infty\! \frac{d w'}{2\pi}\,
    \brh( w') \ln\frac\Lambda{ w'-w-i\ep} -
    \int\limits_{-\infty}^\infty \!dq\,(f(q,u)-\Theta(q))\,
    \bG_{ra}(w-q) \right],
\end{equation}
where $\bar e$ is a RG invariant coupling,
$\Lambda=\mu\exp(\frac{4\pi^2}{e^2})$ is the momentum scale of the
Landau-pole. The derivative of the first term reads
\begin{equation}
  \label{eq:Iw}
  I(w)=\int\limits_{-\infty}^\infty\! \frac{d w'}{2\pi}\,
    \brh( w') \ln\frac\Lambda{ w'-w-i\ep},\quad 
  I'(w)=-\int\limits_{-\infty}^\infty\! \frac{d w'}{2\pi}\,
  \frac{\brh( w')}{w- w'+i\ep} = -\bG_{ra}(w).
\end{equation}
The imaginary part of $I(w)$ term is zero for $w<0$, moreover for
$w=0$ it is negative (at least for large $\Lambda$), while for
$w\to-\infty$ it is positive. So there exists a value $w=-M$ for which
it is zero. Then we can write:
\begin{equation}
  I(w) =-\int\limits_{-M}^w \!dq\, \bG_{ra}(q).
\end{equation}
The scale $M$ replaces the scale $\Lambda$. Assuming that $M\gg T$ we
can change the integration limits to $-M\to M$ in the second part,
too. Then we find
\begin{equation}
  \bz(w) = -\frac{\bar e^2}{4\pi^2} \int \!dq\,f(q,u)\, \bG(w-q),
\end{equation}
where the integral symbol means $\int=\int_{-M}^M$. If it does not cause
problem, we will send $M\to\infty$. The zero temperature part is the
same as in our earlier publication \cite{Jakovac:2011aa}.

Summarizing, the renormalized equation reads now:
\begin{equation}
  \label{eq:renbG}
  w \bG(w) = -\frac{\alpha}{\pi} \int \!dq\,f(q,u)\, \bG(w-q),
\end{equation}
where $f(q,u)$ is given by \eqref{eq:fqu}. Since this equation is
linear, the same will be true for the spectral function (with
different normalization conditions)
\begin{equation}
  \label{eq:renbrh}
  w \brh(w) = -\frac{\alpha}{\pi} \int \!dq\,f(q,u)\, \brh(w-q),
\end{equation}

\subsection{Zero velocity case}
\label{sec:u0}

For $v=0,\, u_0=1$ we find for \eqref{eq:renbrh}
\begin{equation}
  \label{eq:unullcase}
  w \brh(w) = -\frac{\alpha}{\pi} \int \!dq\, (1+n(q)) \brh(w-q).
\end{equation}
By sending the limits of the integration to infinity, we realize that
the right hand side is a convolution. Therefore we change to Fourier
space where it becomes a product, and the left hand side will be
$i\d_ t\brh(t)$. Using the Fourier transform of $1+n(q)$
\begin{equation}
  \pintz w e^{-iw  t} \frac{e^{\beta w}}{e^{\beta w}-1} = 
  \frac{-iT}{2\tanh(\pi  t T)}
\end{equation}
we obtain the differential equation
\begin{equation}
  i\d_ t \brh( t) = \frac{iT\alpha}{\tanh(\pi  t T)}\brh( t).
\end{equation}
This has the following solution:
\begin{equation}\label{u0}
  \brh( t) = \brh_0 \left(\sinh\pi t T \right)^{\alpha/\pi}.
\end{equation}

Before we proceed, we shall discuss this result. First we can easily
recover the $T=0$ result, since for $t\ll \frac1T$ the $\sinh$
function can be approximated linearly, and we get $\brh(t)\sim
t^{\alpha/\pi}$. On the other hand this result is rather weird, it
describes forever increasing correlation instead the physically
sensible loss of correlation. Since this happens also at zero
temperature, this is not an artifact of the finite temperature
calculation. In accordance with Blaizot and Iancu
\cite{IancuBlaizot1,IancuBlaizot2}, we should not consider this
expression as the physical response function. Mathematically we can
argue that we are not in the physically sensible analytic domain, the
time dependent spectral function is not square-integrable for a real
$\alpha$ value, as it should. We must therefore go over to the
physical analytic domain, where the Fourier-transformation is well
defined.

For the analytic continuation we think equation \eqref{u0} valid as
long as it yields sensible formulae, which is the case of
\emph{imaginary} $\alpha$ values. With this assumption the spectral
function in the Fourier space will be an analytic function in
$\alpha$. For real $\alpha$ values the spectral function will be
interpreted as an analytic continuation. We will see that it indeed
provides sensible results.

\begin{figure}[htbp]
  \centering
  \hspace*{-2em}
  \hbox to \hsize{
    \hfill\includegraphics[width=0.4\textwidth]{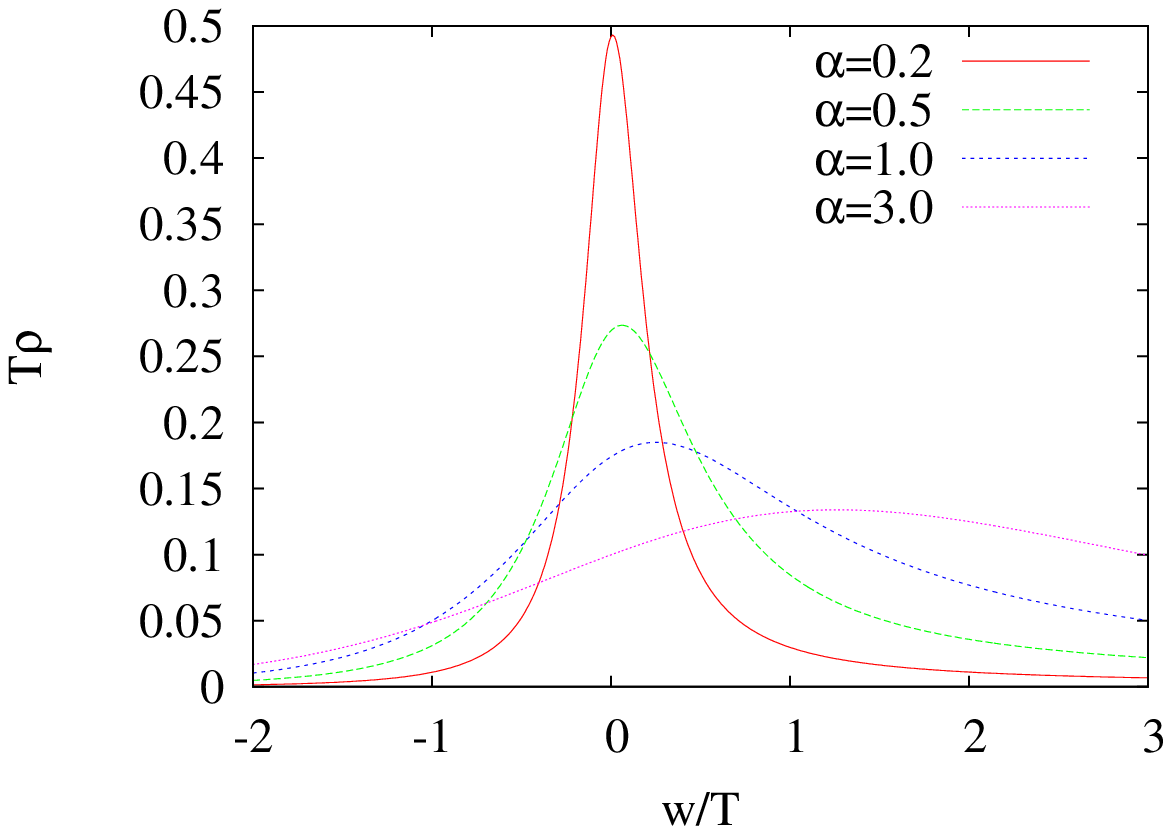}
    \hspace*{2em}
    \includegraphics[width=0.4\textwidth]{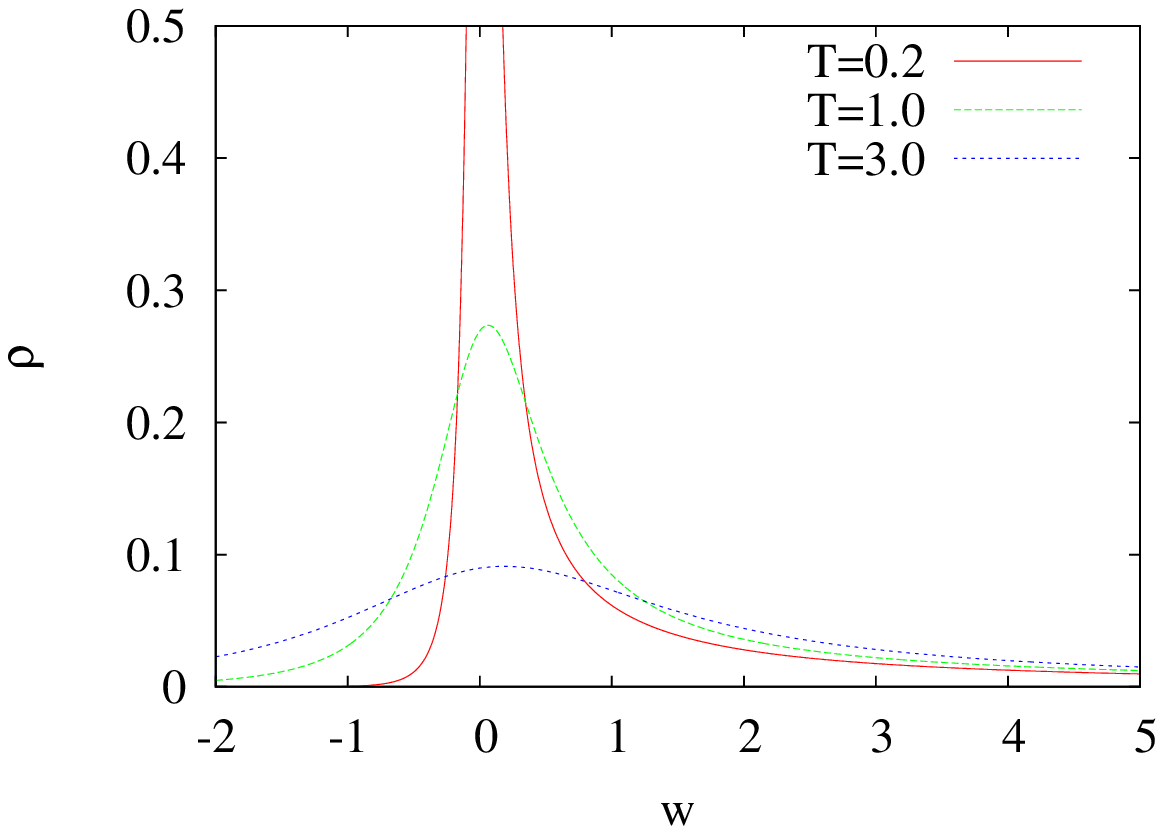}
    \hfill}
  \hbox to \hsize{\hspace*{0.27\textwidth} a.) \hspace*{0.43\textwidth}b.)\hfill}
  \caption{a.) The exact, normalized spectral function at $v=0$. Common
    features are the dominantly exponential decrease for
    $w\to-\infty$, power-law decrease $\sim w^{-1-\alpha/\pi}$ for
    $w\to\infty$ and at the peak a finite curvature $\sim \alpha$. b.)
    Temperature dependence of the spectral function at $v=0$. In the
    limit $T\to0$ it is singular at $w=0$ point.}
  \label{fig:exactsol}
\end{figure}

To perform the inverse Fourier transformation we apply Laplace
transformation. With $s_\pm=\pm iw$ we find
\begin{equation}
  \int\limits_{-\infty}^\infty \!dt\, e^{iw t} \brh(t) =
  \int\limits_0^\infty \!dt\, e^{-s_- t} \brh(t) +
  \int\limits_0^\infty \!dt\, e^{-s_+ t} \brh(-t) = \brh_+(s_-) +
  (-1)^{\alpha/\pi}\brh_+(s_+),
\end{equation}
where
\begin{equation}
  \brh_+(s) = \int\limits_0^\infty \!dt\, e^{-st} \left(\sinh\pi t T
  \right)^{\alpha/\pi} = \frac{\displaystyle \Gamma\left(1+
      \frac\alpha\pi \right)}{\displaystyle 2^{1+\alpha/\pi}\pi T}\; \frac
  {\displaystyle \Gamma\left(\frac {\beta s}{2\pi}
      -\frac{\alpha}{2\pi}\right)} {\displaystyle \Gamma\left(1+\frac
      {\beta s}{2\pi} +\frac{\alpha}{2\pi}\right)}.
\end{equation}
Since the $\Gamma$-function satisfies
$\Gamma(1-z)\Gamma(z)=\pi/{\sin\pi z}$, we can write with $s=\pm iw$:
\begin{equation}
  \left.\frac{\displaystyle \Gamma\left(\frac {\beta s}{2\pi}
      -\frac{\alpha}{2\pi}\right)} {\displaystyle \Gamma\left(1+\frac
      {\beta s}{2\pi} +\frac{\alpha}{2\pi}\right)}\right|_{s=\pm i w} =
  \frac {-\pi} {\displaystyle \left|\Gamma\left(1+\frac{\alpha}{2\pi} +
        i\frac {\beta w}{2\pi}\right)\right|^2
    \sin\left(\frac{\alpha}{2} \mp i \frac {\beta w}{2}\right) }.
\end{equation}
Then we get for the spectral function
\begin{equation}
  \label{eq:exactsol}
  \brh(w) = \frac{\displaystyle N_\alpha \beta \sin\alpha\,e^{\beta w/2}}
  {\displaystyle\cosh(\beta w)-\cos\alpha}\, \frac 1{\displaystyle
    \left|\Gamma\left(1+\frac{\alpha}{2\pi} + i\frac {\beta
          w}{2\pi}\right)\right|^2},
\end{equation}
where $N_\alpha$ is a numerically determined normalization constant
required by the sum rule
\begin{equation}
  \pintz w \brh(w)=1
\end{equation}
to be satisfied. On Fig.~\ref{fig:exactsol} we can see the shape of
the spectral function for different $\alpha$ values and for different
temperatures.

To discuss this result we make the following observations:
\begin{itemize}
\item $\brh(w)$ is a function of $\beta w$ only, which is
  understandable, since there is no other scale in the system which
  could form a dimensionless combination.
\item For $\alpha\to 0$, we find
  \begin{equation}
    \frac{\displaystyle e^{\beta w/2} \sin\alpha}
    {\displaystyle\cosh(\beta w)-\cos\alpha}\to 2\pi\delta(w),
  \end{equation}
  so we recover the free case. It is interesting, that this
  behavior periodically returns for $\alpha=2\pi n$.
\item For large values of $w$ which is equivalent to the small
  temperature case we can use the asymptotic form of the $\Gamma$
  function for complex arguments with large absolute value:
  \begin{equation}
    \Gamma(x) =e^{-x}x^x \left(x^{-1/2} +{\cal O}(x^{-3/2})\right).
  \end{equation}
  Then we find, up to normalization factors
  \begin{equation}
    \brh(\beta w\gg1) \sim \frac{\displaystyle e^{\beta w}}
    {\cosh(\beta w)}\, \frac 1{w^{1+\frac\alpha\pi}}\;
    \stackrel{T\to0}{\longrightarrow}\; \Theta(w) w^{-1-\frac\alpha\pi}.
  \end{equation}
  This is the well-known exact solution by Bloch and Nordsieck at zero
  temperature. Note that the $\Theta$ function came out correctly from
  the formula. At finite but small temperatures, for negative
  arguments we observe exponential decrease.

  This form also shows how at zero temperature we obtain zero wave
  function renormalization factor. The normalization factor
  (c.f. \eqref{eq:exactsol}) is proportional to $\beta$, while the
  asymptotic form is $(\beta w)^{-1-\frac\alpha\pi}$. Then approaching
  zero temperature we obtain $T^{\frac\alpha\pi}
  w^{-1-\frac\alpha\pi}$, which means a renormalization factor
  vanishing as $\sim T^{\frac\alpha\pi}$ for $T\to0$.

\item Now let us consider the $w\to0$ limit, i.e. the vicinity of the
  mass shell. We can expand $\brh$ into power series
  \begin{equation}
     \brh(w) = \frac{4\brh(0)C T^2}{\displaystyle(w-CT)^2 +
       \left(\frac4C-1\right) C^2T^2 +{\cal O}(w^3)},
  \end{equation}
  where
  \begin{equation}
    \frac1C=\frac12 + \frac{2}{1-\cos\alpha} + \frac1{\pi^2}
    \Psi^2(1+\frac{\alpha}{2\pi}),
  \end{equation}
  and $\Psi(a)$ is the digamma function. The maximum of this function
  is at $CT$, the width is $CT \sqrt{4C^{-1}-1}$. Since, however, the
  function is not symmetric, these parameters cannot be interpreted as
  a thermal mass and thermal width. For that we need to examine the
  real time dependence.

\item For the real time dependence we use the fact that, according to
  \eqref{eq:exactsol}, $\rh(p) = \beta f_0(\beta (p_0-m))$, which
  means that $\rh(t) = e^{-i m t} \tilde f_0(Tt)$. Omitting the
  oscillating phase (i.e. if we consider the envelope of $\rh(t)$),
  we recover the Fourier transform of $f_0$.

  The real time dependence obtained from the inverse Fourier
  transformation of \eqref{eq:exactsol} differs from \eqref{u0}. This
  is because we performed an analytic continuation to the physically
  sensible analytic domain. The numerical inverse Fourier
  transform of the normalized spectral function (and, because
  $iG_{ra}(t)=\Theta(t)\rh(t)$, for $t>0$ this is also the real time
  dependence of the retarded Green's function) can be seen on
  Fig.~\ref{fig:timedep}.
  \begin{figure}[htbp]
    \centering
    \includegraphics[width=0.4\textwidth]{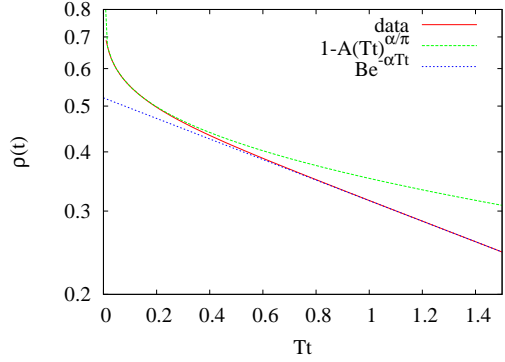}
    \caption{Time dependence of the (envelope of the) retarded Green's
      function (or, equivalently, the spectral function) for $v=0$ at
      $\alpha=0.5$ on a logarithmic y-scale. For small times we find
      $1-A(Tt)^{\alpha/\pi}$, corresponding to the zero temperature time
      dependence. For large times it turns into an exponential
      $\exp(-\alpha Tt)$ damping.}
    \label{fig:timedep}
  \end{figure}

  At small times we expect to recover the zero temperature
  result. Indeed, we observe $\brh(t) =
  (1-A(Tt)^{\alpha/\pi})e^{-imt}$ asymptotic form (for $\alpha=0.5$
  this is valid up to $Tt<0.4$), the power law time dependence is
  characteristic to the zero temperature result. At $t=0$ the value of
  the spectral function is 1, this is because of normalization. Note
  however, that naively at zero temperature we would obtain $\sim
  t^{\alpha/\pi}e^{-imt}$ time dependence, describing growth of
  correlation and violating the normalization condition. Interpreting
  the zero temperature result as $T\to 0$ limit, we could cure this
  apparent inconsistency of the model. At strictly $T=0$ we get back
  the physically sensible oscillating solution $\rh(t) = e^{-imt}$.

  For large times (for $tT>1$) the time dependence is $\sim e^{-\alpha
    Tt}$, which agrees with
  \cite{IancuBlaizot1,IancuBlaizot2}. Comparing it to \eqref{u0} we
  see that instead of an exponential rise we found an exponential
  decay, but with the same coefficient. This can be understood by
  noting that if we have a pole at $w=w_0$ in the momentum space,
  meaning $e^{-iw_0 t}$ exponential time dependence, this pole is
  present in the spectral function in position $w_0^*$, too. The
  physical retarded Green's function can have poles in the lower half
  plane, therefore we find in our case only the $w_0=-i\alpha T$ pole,
  giving exponential damping.
\end{itemize}

\begin{figure}[htbp]
  \centering
  \includegraphics[width=0.4\textwidth]{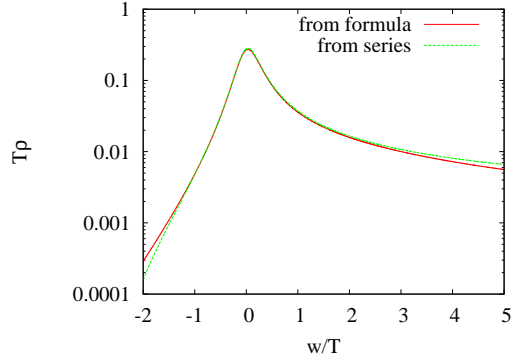}
  \caption{Comparison of the logarithm spectral function at $\alpha=0.5$
    calculated from eq. \eqref{eq:exactsol} and from
    eq. \eqref{eq:spectseries}. The two results agree well.}
  \label{fig:u0spect}
\end{figure}

For the justification of the analytic continuation we also used a
different method. We expanded the $t$-dependent result \eqref{u0} into
power series using
\begin{equation}
  (\sinh x)^{\alpha/\pi} = \left(\frac12\right)^{\frac{\alpha}{\pi}}
  \sum\limits_{k=0}^{\infty} \left[ \Theta(x) (-1)^k {\frac{\alpha}{\pi}
      \choose k} e^{x(\frac{\alpha}{\pi}-k)} e^{-xk} + \Theta(-x)
    (-1)^{\frac{\alpha}{\pi}-k} {\frac{\alpha}{\pi} \choose k}
    e^{-x(\frac{\alpha}{\pi}-k)} e^{xk}\right]
\end{equation}
Now the inverse Fourier transformation acts on a pure exponential
function. We use the formula
\begin{equation}
  \int\limits_0^\infty\!dt\,e^{\pm iw t - s t} =
  \frac1{s\mp iw}
\end{equation}
which is true, of course, if $s>0$, but this is the formula for the
analytic continuation, too. Then the result of the Fourier
transformation, with appropriate normalization to ensure reality of
$\rh$:
\begin{equation}
  \rh(w) \sim \sum\limits_{k=0}^{\infty} (-1)^k
  {\frac{\alpha}{\pi} \choose k} \left[
    \frac{(-1)^{-\alpha/2\pi}}{s_k+iw} +
    \frac{(-1)^{\alpha/2\pi}}{s_k-iw}\right], 
\end{equation}
where $s_k=\pi T(2k-\alpha/\pi)$. Using the $(-1)^{\alpha/2\pi} =
\cos\frac\alpha2 + i \sin\frac\alpha2$ definition we find after a
simple calculation
\begin{equation}
  \label{eq:spectseries}
  \rh(w) \sim \sum\limits_{k=0}^{\infty} (-1)^k
  {\frac{\alpha}{\pi} \choose k} \frac{s_k (1+\cos\alpha) -
    w\sin\alpha}{s_k^2+w^2},\qquad s_k=\pi
  T(2k-\frac\alpha\pi).
\end{equation}
The sum converges fast, and we can compare the result of the two
calculations on Fig.~\ref{fig:u0spect}.
We can see that the two methods of analytic continuation yield
consistent result in the central peak regime. To understand the small
deviations at the edges, we remark that if $\alpha$ is real then
$\brh(t=0)=0$ (c.f. \eqref{u0}). In Fourier space this means $\int dw
\brh(w)=0$, therefore it can not be positive for all momenta. Using
the second method, the position where the spectral function turns into
negative values is, fortunately, at large $|w/T|$ values, therefore
the peak is unaffected. The only precursor of the sign changing is the
slight decrease at the edges of the plot. In our first method we
started from imaginary $\alpha$ values, where
$\lim\limits_{t\to0}\brh(t)=\brh_0\neq0$, then the normalization does
not require negative values for $\brh(w)$.

\subsection{Finite velocity case}
\label{sec:ufin}

If $v\neq0$ we find for eq. \eqref{eq:renbrh} using \eqref{eq:fqu} the
following formula
\begin{equation}
  \label{eq:finu0}
  w \brh(w) = -\frac{\alpha}{\pi}  \frac{u_0(1-v^2)}{2v}\!\!\!
  \int\limits_{u_0(1-v)}^{u_0(1+v)}\!\!\!\frac{ds}{s^2}\, \int \!dq\,
  (1+n(\frac qs)) \,\brh(w-q).
\end{equation}
The right hand side is again a convolution, and formally we can use
the same method as in the $v=0$ case. We find
\begin{equation}
  \d_ t \brh( t) = \frac{\alpha}{\pi}\frac{u_0(1-v^2)}{2v}\!\!\!
  \int\limits_{u_0(1-v)}^{u_0(1+v)} \!\!\! \frac{ds}{s^2}\,
  \frac{Ts}{\tanh(\pi  t Ts)}\brh( t)
\end{equation}
which has the solution
\begin{equation}
  \label{u1}
  {\brh( t)} = \brh(0)\exp\left[ \frac{\alpha}{\pi}
    \frac{u_0(1-v^2)}{2v}\!\!\! \int\limits_{u_0(1-v)}^{u_0(1+v)}
    \!\!\! \frac{ds}{s^2}\, \ln(\sinh\pi t Ts)\right].
\end{equation}
We cannot perform analytically neither the integral, nor its Fourier
transform. But we can determine many features by investigating the
$t\to0$ and $t\to\infty$ limits.

\begin{subsubsection}{The limit $t\rightarrow 0 $}

Since $\lim\limits_{t\rightarrow 0} \sinh\pi t Ts=\pi tTs$:
\begin{equation}
  \lim_{t\rightarrow 0}\frac{u_0(1-v^2)}{2v}\!\!\!
  \int\limits_{u_0(1-v)}^{u_0(1+v)} \!\!\! \frac{ds}{s^2}\,
  \ln(\sinh\pi t Ts) = \frac{u_0(1-v^2)}{2v}\!\!\!
  \int\limits_{u_0(1-v)}^{u_0(1+v)} \!\!\!\frac{ds}{s}\, \ln
  \pi tTs= \ln \pi Tt+\text{const.},
\end{equation}
where the constant comes from the integral of $s^{-2}\ln s$. Being a
finite quantity, it goes into the normalization. After exponentiation
we find
\begin{equation}
  \brh( t)\sim (Tt)^{\alpha/\pi},
\end{equation}
which is the zero temperature result. So, as we expected the short
time or large frequency regime reproduces the zero temperature case,
and thus it is velocity-independent.

\end{subsubsection}

\begin{subsubsection}{The limit $t\rightarrow \infty $}

Here the $\sinh$ can be approximated by the exponential, and so
\begin{equation}
  \lim\limits_{t\rightarrow \infty}\ln \sinh\pi Tts =\pi Tts-\ln2 
\end{equation}
The $\ln2$ yields a constant factor which goes into the
normalization. The rest gives, including the prefactors
\begin{equation}
  \alpha T t \frac{u_0(1-v^2)}{2v}\!\!\! \int\limits_{u_0(1-v)}^{u_0(1+v)}
  \!\!\!\frac{ds}s  =  \alpha_{eff}(u) Tt,
\end{equation}
where
\begin{equation}
  \label{eq:alphaeff}
  \alpha_{eff}(u)=\alpha \frac{u_0(1-v^2)}{2v} \ln
  \left(\frac{1+v}{1-v}\right).
\end{equation}
From this form we obtain for the spectral function in the asymptotic limit:
\begin{equation}
  \brh( t)=Ce^{\alpha_{eff}(u)Tt}.
\end{equation}
We can easily check that $\lim\limits_{v\to0} \alpha_{eff}(u)=\alpha$.
Therefore the $v\to0$ limit is analytic.

Since in the asymptotic time regime we simply get the substitution
rule $\alpha\to\alpha_{eff}(u)$ as compared to the $v=0$ case, the
analysis of the vicinity of the peak of the spectral function, and the
large time dependence will remain valid in the finite velocity case,
too, with a modified value of the coupling. In particular, since
$\alpha_{eff}(u)<\alpha$, we obtain a smaller damping, larger lifetime
for $v>0$ cases. Physically this property is the consequence of the
decreasing cross section at larger energies.
\end{subsubsection}

\begin{subsubsection}{Solution for $t\in(0,\infty)$}

  For intermediate times we could not work out analytically the
  integral. Nevertheless, we have a well-controlled numerical method
  to find the spectral function, once the analytic behavior for large
  $t$ is identified. We express the wanted $\brh_u(t)$ as a product of
  the known $\brh_{u=0}(t;\alpha_{eff})$ and a correction factor
\begin{equation}
  \brh(t)\sim Z(t)\brh_{u=0}(t;\alpha_{eff}),
\end{equation}
where
\begin{equation}
  Z(t)\equiv \frac{\displaystyle \exp\left[\frac{\alpha
          u_0(1-v^2)}{2\pi v}\!\!\! \int\limits_{u_0(1-v)}^{u_0(1+v)}
        \!\!\! \frac{ds}{s^2} \ln(\sinh\pi t Ts)\right]
    }{\displaystyle (\sinh\pi T t)^{\frac{\alpha_{eff}(u)}{\pi}}}.
\end{equation}
After a short algebra we find
\begin{equation}
  Z(t)=\exp\left\{  \frac{\alpha u_0(1-v^2)}{2\pi v}\!\!\!
    \int\limits_{u_0(1-v)}^{u_0(1+v)} \!\!\!\frac{ds}{s^2}
    \ln\frac{\sinh\pi Tts}{(\sinh\pi Tt)^s}\right\}.  
\end{equation}
The so-defined ratio is symmetric $Z(t)=Z(-t)$. For small $t$
arguments it behaves as $Z(t)\sim
(Tt)^{\frac{\alpha-\alpha_{eff}(u)}{\pi}}$ and, since
$\alpha\geqq\alpha_{eff}(u)$, we also know $Z(t=0)=0$. At large $t$ we
find $\lim\limits_{t\rightarrow \infty}Z(t)=1$. We can determine it
numerically, for a specific $v$ it can be seen on Fig.~\ref{fig:Zfunct}
\begin{figure}[htbp]
  \centering
  \includegraphics[width=0.4\textwidth]{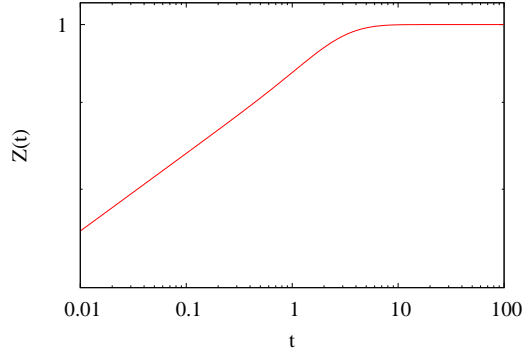}
  \caption{The $Z(t)$ function on a logarithmic plot. For small times
    it is a power, for larger times it flattens out.}
  \label{fig:Zfunct}
\end{figure}

We can numerically Fourier transform $Z(t)$, and perform a convolution
in the Fourier space with the $\brh_{v=0}(w)$ function
\eqref{u0}. This ensures that we use the same analytic continuation
for the different velocity cases. As a result we obtain
Fig.~\ref{fig:vdep}.
\begin{figure}[htbp]
  \centering
  \hbox to \hsize{
    \hfill\includegraphics[width=0.4\textwidth]{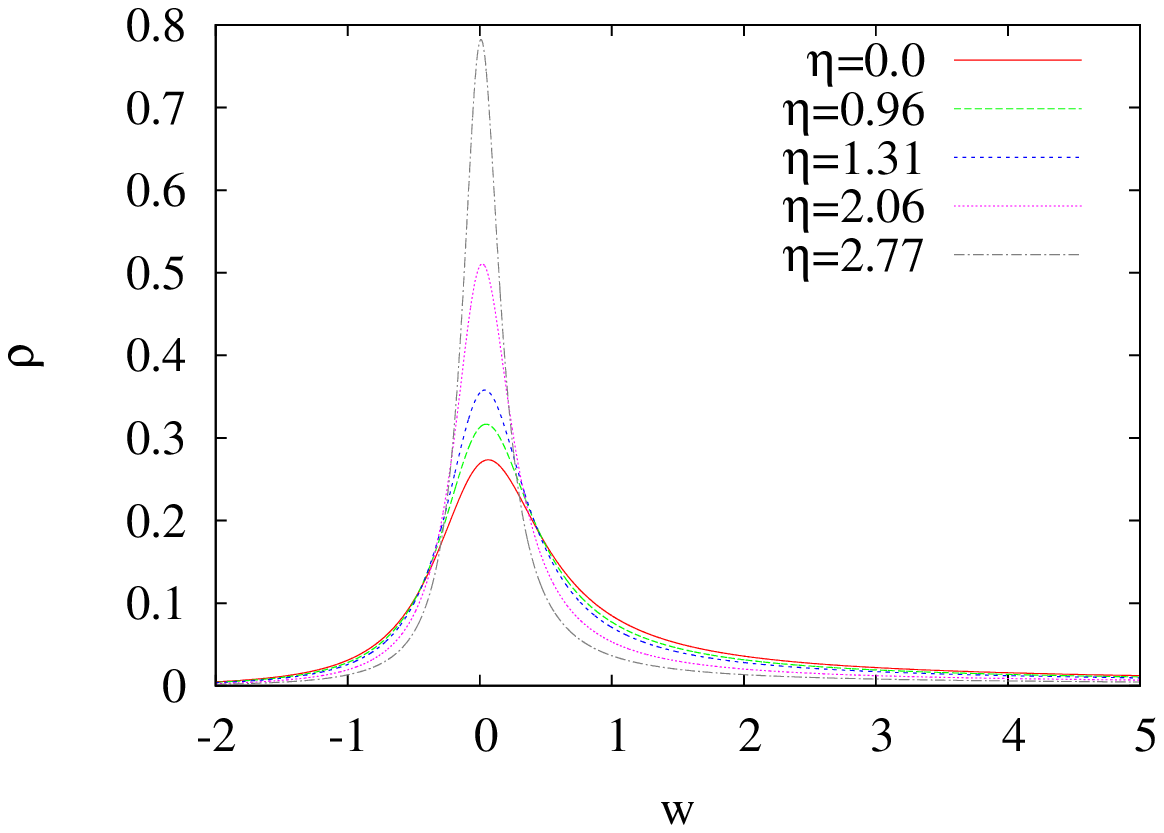}
    \hspace*{2em}
    \includegraphics[width=0.4\textwidth]{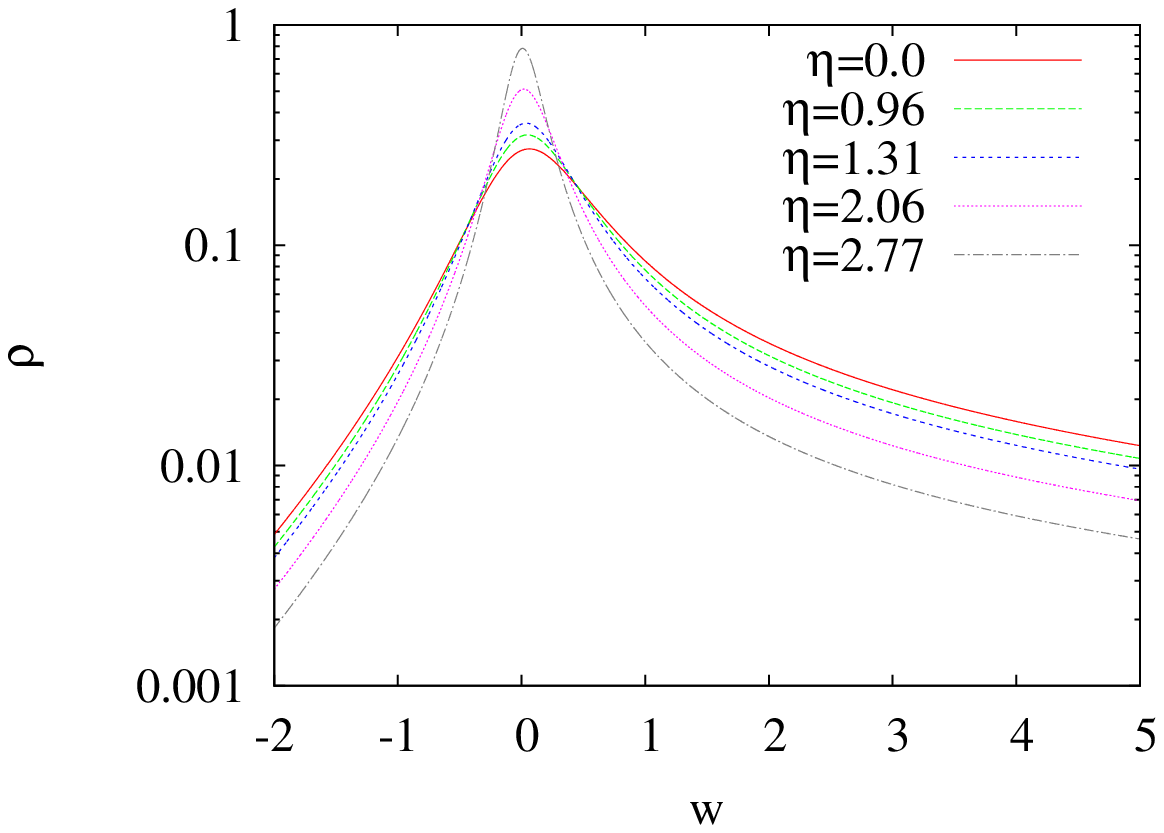}
    \hfill}
  \hbox to \hsize{\hspace*{0.27\textwidth} a.) 
    \hspace*{0.43\textwidth}b.)\hfill}
  \caption{Velocity dependence of the spectral function for
    $\alpha=0.5$. The $\eta$ values are rapidities, $v=\tanh\eta$. a.)
  is a linear-linear plot to show that the peak region becomes more
  and more peaked with increasing $\eta$ ($v$). The log-log plot
  demonstrates that the asymptotics remain the same.}
  \label{fig:vdep}
\end{figure}
We can observe that the peak becomes narrower for larger velocities,
corresponding to the decreasing $\alpha_{eff}$ value. At large
momentum the asymptotics is the same for all velocities (for a given
$\alpha$), because the zero temperature result is insensitive
to the value of $v$.

Here again we can work out the real time dependence. We now write
$\rh_u(p)= u_0 \beta f_u(\beta(p_0 u_0-\mathbf{pu}-m))$ and find
\begin{equation}
  \rh_u(t)= e^{-i \mathbf{vp}t - imt/u_0} \tilde f_u(\frac{tT}{u_0}).
\end{equation}
The result of the numerical inverse Fourier transform can be seen on
Fig.~\ref{fig:timedepv}.
\begin{figure}[htbp]
  \centering
  \includegraphics[width=0.4\textwidth]{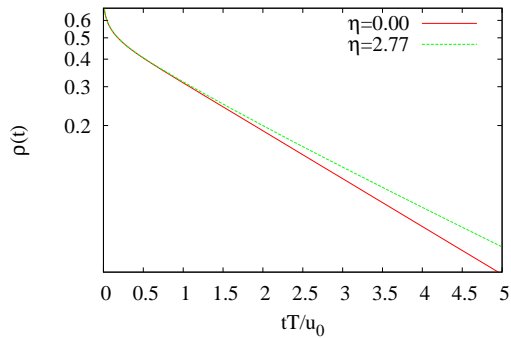}
  \caption{Comparison of the real time dependence of the retarded
    Green's function (or, equivalently, the spectral function) for zero
    and finite velocity at $\alpha=0.5$ on logarithmic y scale. The
    small time behavior does not change, at large times the
    exponential damping turns to $\exp(-\alpha_{eff}(u) Tt)$.}
  \label{fig:timedepv}
\end{figure}
At small times the spectral function (retarded Green's function) is
velocity independent, this is the zero temperature asymptotics. For
large times (for $Tt>1$) the time dependence turns into $\sim
e^{-\alpha_{eff}(u) Tt}$.
\end{subsubsection}

\subsection{Discussion of earlier results}
\label{sec:disc}

We can compare our results to the earlier results in the literature.
The long time asymptotics of the finite temperature solution of the
Bloch-Nordsieck model was already discussed in
\cite{IancuBlaizot1,IancuBlaizot2}. They followed a different,
functional approach. Still, the two methods lead to the same
intermediate result. Neglecting renormalization effects (which is
treated later in Ref.\ \cite{IancuBlaizot1,IancuBlaizot2}), our
eq. \eqref{eq:master} together with \eqref{eq:JJ2} yields, using the
notation $w=up-m$
\begin{equation}
  w \G_{ra}(w) = e^2U^2 \pint4k \frac1{uk}\, (1+n(k_0)) \bar\rh(k)\,
  \bG_{ra}(w-uk),
\end{equation}
where $\bar\rh(k)=\frac{2\pi}{2k} (\delta(k_0-k) - \delta(k_0+k))$ is
the photon spectral function. After Fourier-transformation we find
exponentiation of the real time contributions
\begin{equation}
  \G_{ra}(t) = \G_{ra}(t=0) e^{F(t)},
\end{equation}
where
\begin{equation}
  F(t) = -e^2U^2 \pint4k (1+n(k_0)) \bar\rh(k)\,\frac{1-e^{-iukt}}
  {(uk)^2} = -it\Phi_0 + it\Phi(t) + \ln\Delta(t),
\end{equation}
where $\Phi_0$, $\Phi(t)$ and $\ln\Delta(t)$ are real quantities,
corresponding to the notation of \cite{IancuBlaizot2}. Using $U^2=1$
this means
\begin{eqnarray}
  && \Phi(t) = -e^2 \pint4k \frac{\bar\rh(k)}{uk} \left[ 1- \frac{\sin
      ukt}{ukt}\right] \nn
  && \Phi_0 = -e^2 \pint4k \frac{\bar\rh(k)}{uk} \nn
  && \Delta = \exp\left\{-e^2 \pint4k n(k_0) \bar\rh(k)\,\frac{1-\cos ukt}
  {(uk)^2}\right\}.
\end{eqnarray}
These expressions agree with the equations (2.24) and (2.25) of
\cite{IancuBlaizot2} (the constant phase $\Phi_0$ has no physical
meaning).

The analysis of this formula, however, differs in our case and in
\cite{IancuBlaizot1,IancuBlaizot2}. We strictly restrict ourselves to
the original Bloch-Nordsieck model, and used the free photon spectral
function. In Ref.\ \cite{IancuBlaizot1,IancuBlaizot2} the authors used
HTL-improved photon spectral function (cf. their eq. (3.1) and
(3.2)). As it turns out, the most important contribution comes from
the small frequency limit of the continuum (Landau damping) part. This
explains why the asymptotic time behavior differs in our case and in
the case of Ref.\ \cite{IancuBlaizot1,IancuBlaizot2} (
$\exp(-\alpha_{eff}(v)Tt)$ vs. $\exp(-C t \log t)$ ).

In Ref.\ \cite{Fried:2008tb} Fried \emph{et.\@ al.\@} use again a
different formalism. Since they examine a different physical
situation, the comparison is much more difficult. What is clear,
however, that they also use the original version of the model, and
also find exponentially damping solution.

It is very interesting that in \cite{Wang:2000via} the authors found
the same $\exp(-\alpha T t \log t)$ like solution as was the case in
Ref.\ \cite{IancuBlaizot1,IancuBlaizot2}, although with a different
line of thought. They use the dynamical renormalization group idea
\cite{Boyanovsky:1999cy}, where the secular terms are melted into
finite time dependence of the renormalized parameters. Clearly they
cannot consider all photonic diagrams, just those which contribute to
the renormalization group (RG) equations. The logarithmic enhancement
of the damping there can be interpreted physically as an eternally
growing cross section of the incoming hard particle which collects
more and more soft photons around itself. The analysis of the pure
Bloch-Nordsieck model results in a finite damping, which means that in
this model the initial growth of the cross section eventually stops,
the soft photon cloud saturates. The physical interpretation of the
saturation probably is that the multi-photon contributions arriving
from different spacetime points become incoherent.

The two scenarios, one with ever growing photon cloud, the other with
saturation, are both approximations of the real QED (RG, HTL and free
photon approximations, respectively). The question that which one is
finally manifested in QED, can be answered only after a full analysis
of the complete QED where all these effects are present.

\section{Conclusions}
\label{sec:concl}

In this paper we studied the Bloch-Nordsieck model at finite
temperature, in particular we studied the fermionic spectral
function. We used the strategy introduced in \cite{Jakovac:2011aa}
which is based on the Dyson-Schwinger equations, where the infinite
hierarchy is closed by using the Ward identities for the vertex
function. We worked out the corresponding equations at finite
temperature in the real time formalism, and solved them. This
procedure is exact in the Bloch-Nordsieck model.

At zero velocity we were able to obtain fully analytic results for the
spectral function. For large momenta and/or zero temperature this
formula agrees with the zero temperature result. At finite temperature
there appears an asymmetric peak which decreases exponentially below
the mass shell ($u^\mu p_\mu <m$) and as a power law above the mass
shell ($u^\mu p_\mu >m$). 

We also worked out the real time dependence which has two
characteristic regime. For small times, starting from its initial
value, it behaves as a power law $\brh(t) =
(1-A(Tt)^{\alpha/\pi})e^{-imt}$, where $A$ depends on $\alpha$. The
naive zero temperature calculation yields $\sim
t^{\alpha/\pi}e^{-imt}$ time dependence which is not normalizable and
corresponds to a physically hardly interpretable forever growing
retarded response function. With the finite temperature as a regulator,
we could interpret the zero temperature result, and we got a
physically sensible purely oscillating response function.

For large times we find exponential damping. The damping rate is
$\alpha_{eff}(u)T$, where the effective coupling $\alpha_{eff}(u)$ is
given in \eqref{eq:alphaeff}. The damping is smaller, the lifetime is
longer for larger velocities, which physically can be interpreted as
the consequence of decreasing cross sections. We remark that the
damping in the pure Bloch-Nordsieck model differs from the one with
HTL-improved photon propagator; in this latter case one finds a
faster-than-exponential damping with an exponent $\sim-t\ln t$.

We expect that the method we worked out for the Bloch-Nordsieck model
can be applied, as an approximation scheme, also for the full
QED. Hopefully the renormalizability of the resummation will remain
true, too.

\begin{acknowledgments}
  The authors acknowledge useful discussions with K. Homma,
  M. Horv\'ath, G. Mark\'o, A. Patk\'os, U. Reinosa and
  Zs. Sz\'ep. The project was supported by the Hungarian National Fund
  OTKA-K68108 and OTKA-K104292 and the New Sz\'echenyi Plan (Project
  no. T\'AMOP-4.2.2.B-10/1--2010-0009). Feynman diagrams were drawn by
  JaxoDraw program.
\end{acknowledgments}

\end{document}